\documentclass[twocolumn,twoside,superscriptaddress,prl,aps,showpacs]{revtex4}
\usepackage{amsmath,mathrsfs,amsbsy,color,graphicx,bm,amsthm,amsfonts}
\usepackage{units}
\usepackage{bbm}
\usepackage{times}
\usepackage{dcolumn}
\usepackage{mathrsfs}% Align table columns on decimal point
\usepackage{amsmath,amssymb,epsfig,float}

\begin{document}

\title{The influence of the Earth's curved spacetime on Gaussian quantum coherence}
\author{Tonghua Liu$^{1}$, Shumin Wu$^{2}$, Shuo Cao \footnote{Email: caoshuo@bnu.edu.cn}}
\affiliation{ Department of Astronomy, Beijing Normal University, Beijing ,100875, China\\
$^2$Department of Physics, and Collaborative Innovation Center for Quantum Effects \\
and Applications,
 Hunan Normal University, Changsha, Hunan 410081, China}

\iffalse
\author{Tong-Hua Liu$^1$}
\author{Shuo Cao$^1$}
\email{hszeng@hunnu.edu.cn}
\author{Shu-Min Wu$^2$}
\author{Hao-Sheng Zeng$^2$}
\affiliation{ $^1$ Department of Physics, Synergetic  Innovation Center for Quantum Effects, and Key Laboratory of Low
Dimensional Quantum Structuresand Quantum Control \\
 of Ministry of Education,
 Hunan Normal University, Changsha, Hunan 410081, China\\
 $^2$ Department of Astronomy, Beijing Normal University, Beijing ,100875, China.}\fi

% \baselineskip=0.65 cm

\begin{abstract}
Light wave-packets propagating from the Earth to satellites will be deformed by the curved background spacetime of the Earth, thus influencing the quantum state of light. We show that Gaussian coherence of photon pairs, which are initially prepared in a two-mode squeezed state, is affected by the curved spacetime background of the Earth.  We demonstrate that quantum coherence of the state increases for a specific range of height $h$ and then  gradually approaches a finite value with further increasing height of the satellite's orbit in Kerr spacetime, because special relativistic effect are involved.  Meanwhile, we find that Gaussian coherence increases
with the increase of Gaussian bandwidth parameter, but the
Gaussian coherence decreases with the growth of the peak
frequency. In addition, we also find that total gravitational frequency shift causes changes of Gaussian coherence less than $\% 1$ and different initial peak frequencies also can effect rate of change with the satellite height in geostationary Earth orbits.
\end{abstract}
\vspace*{0.5cm}
 \pacs{XX.XX.XX No PACS code given }
\maketitle

%----------------------------------------------------------------------------------------------------------------------------------------------------------------------------------------------------------------------------%
\section{Introduction}
%----------------------------------------------------------------------------------------------------------------------------------------------------------------------------------------------------------------------------%

The coherent superposition of states
stands as one of the characteristic features that mark the
departure of quantum mechanics from the classical realm \cite{E.P.C}. Unlike quantum entanglement, discord, quantum coherence can exist in single systems, and can be achieved efficiently or impossible by classical methods. Quantum coherence constitutes
a powerful resource for quantum metrology \cite{EPC1, EPC2}
and entanglement creation \cite{EPC3, EPC4} and is the fundamental physical explanation of a series of intriguing phenomena
in quantum optics \cite{EPC5, EPC6, EPC7,EPC8} and quantum information \cite{EPC9}. Viewing quantum coherence as resources is crucial for developing new quantum technologies. Recently, the necessary criteria for valid quantifiers of
coherence and the rigorous characterizations
of coherence in the framework of resource theories have been put forward
in \cite{EPC10, EPC11, EPC12, EPC13}. Whereafter, a subsequent
stream of works has identified coherence measures for both
theoretical and experimental purposes \cite{HMW,Skrzypczyk, Kocsis, Adesso2015, steering3, steering4,MWZZ,steering5,Walborn,steering2,Bowles,VHTE}.
However, more attention has been given to quantum coherence without relativity effects, only little is known about behaviors of quantum coherence in a relativistic setting or curved spacetime background. Recently, quantum coherence had been studied in the dynamical Casimir effect in \cite{DSSC}. In addition, it has been given to the dynamics of quantum coherence under the accelerated motions \cite{TBM}.

Since realistic quantum systems  always exhibit gravitational and relativistic features,
quantum system cannot be prepared and transmitted in a curved
spacetime without any gravitational and relativistic effects,
the study of quantum coherence in a relativistic framework is necessary.  Understand the influence of gravitational effects on the coherence quantum resource has practical and  fundamental significance in realistic world when the parties involved are located at large distances in the curved space time \cite{DSSC, TBM}. The
curved background spacetime of the Earth  affects the running of quantum clocks, is employed as witnesses of general
relativistic proper time in laser interferometric \cite{Zych}, and  influences the implementation of quantum metrology \cite{MADE, MADE2} in satellite-based setups has been proposed in Refs \cite{Alclock, wangsyn}. Kish and Ralph found that there would be inevitable losses of quantum resources
in the estimation of the Schwarzschild radius \cite{SPK}. Furthermore Satellite-based quantum steering under the influence of spacetime curvature of the Earth had been proposed in Ref. \cite{TQJJH}.

In this work, we present a quantitative investigation of Gaussian
quantum coherence for correlated photon pairs which are initially prepared in a two-mode squeezed state under the
curved background spacetime of the Earth. We assume that one of entangled photons stay at Earth's surface and the other propagates to the satellite. During this propagation, the photons' wave-packet will be deformed by the curved background spacetime of the Earth, and these deformations effects on the quantum state of the photons can be modeled as a lossy quantum channel \cite{MANI}.  We quantitative calculate how much the losses of Gaussian quantum coherence  and  also discuss the behaviors of Gaussian coherence under the gravity of the Earth.

This work is organized as follows. In section II, we introduce the quantum field theory of a massless uncharged bosonic field which propagates from the Earth to a satellite. In section III, we briefly introduce the definition of the measurement of a bipartite Gaussian quantum coherence. In section IV, we show a scheme to test  large distance quantum coherence between the Earth and satellites and study the behaviors of quantum coherence in the curved spacetime.  The last section is devoted to a brief summary. Throughout the whole paper we employ natural units $G = c =\hbar= 1$.

\section{Light wave-packets propagating on Earth's space-time \label{tools}}
%------------------------------------------------------------------------------------------------------------------------------------------------------------------------------------------------%
In this section we will a briefly introduce about the propagation of photons from the Earth to satellites under the influence of the Earth's gravitational field \cite{DEBT}. The Earth's spacetime can be approximately described by the Kerr metric \cite{Visser}. For the sake of simplicity, our work will be constrained to the equatorial plane $\theta=\frac{\pi}{2}$. The reduced metric in Boyer-Lindquist coordinates $(t,r,\phi)$  reads \cite{Visser}
\begin{align}\label{metric}
ds^2=&\, -\Big(1-\frac{2M}{r} \Big)dt^2+\frac{1}{\Delta}dr^2 \nonumber \\
&\,+\Big(r^2+a^2+\frac{2Ma^2}{r}\Big) d\phi^2 - \frac{4Ma}{r} dt \, d\phi, \\
\Delta=&\,1-\frac{2M}{r}+\frac{a^2}{r^2},
\end{align}
where $M$, $r$, $J$, $a=\frac{J}{M}$ are the mass, radius, angular momentum and Kerr parameter of the Earth, respectively.

In order to better describe the propagation of wave-packets from a
source on Earth to a receiver satellite situated at a fixed distance from the source this process, we assume Alice on Earth's surface (i.e. $r_A=r_E$) and Bob who is on a satellite at a radius $r_B$.
A photon is sent from Alice to Bob at time $\tau_A$, Bob will receive this photon at $\tau_B=\Delta\tau+\sqrt{f(r_B)/f(r_A)}\tau_A$ in his own reference frame, where $f(r_A)=1-\frac{r_S}{r_A}$ and  $f(r_B)=1-\frac{r_S}{r_B}$. Here $r_S=2M$ is the Schwarzschild radius of
the Earth and $\Delta\tau$ is the propagation time of the light from the Earth to the satellite by taking curved effects of the Earth into account. Realistic photon sources do not produce monochromatic photons, a photon can be modeled by a wave packet of excitations of a massless bosonic field with a distribution $F^{(K)}_{\Omega_{K,0}}$ of mode frequency $\Omega_{K}$ and peaked at $\Omega_{K,0}$ \cite{ULMQ,TGDT}, where $K=A,B$ denote the modes in Alice's or Bob's reference frames, respectively.  The annihilation operator for the photon for an observer infinitely far from Alice or Bob, takes the form
\begin{equation}
\hat{a}_{\Omega_{K,0}}(t_K)=\int_0^{+\infty}d\Omega_K e^{-i\Omega_K t_K}F^{(K)}_{\Omega_{K,0}}(\Omega_K)\hat{a}_{\Omega_K}.
\label{wave}
\end{equation}
Alice's and Bob's operators in Eq. (\ref{wave}) can be used to describe the same optical mode in  different altitudes. The photon's creation $\hat{a}^{\dagger}_{\Omega_{K,0}}$ and annihilation operators $\hat{a}_{\Omega_{K,0}}$ satisfy the canonical equal time bosonic commutation $[\hat{a}_{\Omega_{K,0}}(t),\hat{a}^{\dagger}_{\Omega_{K,0}}(t)]=1$
relations when  the frequency distribution $F^{(K)}(\Omega)$ is normalized, that is $\int_{\Omega>0}|F^{(K)}(\Omega)|^2=1$. This distribution naturally models a photon which is a wave packet of the electromagnetic field that propagates and is localized in space and time.

Considering the Earth's gravitational field between Alice and Bob,
the wave packet received by Bob is modified when Alice sent a wave packet of the photon.  The relation between $\hat{a}_{\Omega_A}$ and $\hat{a}_{\Omega_B}$ was discussed in \cite{DEBT,DEBA,wangsyn}, and can be used to calculate the relation between the frequency distributions $F^{(K)}_{\Omega_{K,0}}$ of the photons before and after the propagation
\begin{eqnarray}
F^{(B)}_{\Omega_{B,0}}(\Omega_B)=\sqrt[4]{\frac{f(r_B)}{f(r_A)}}F^{(A)}_{\Omega_{A,0}}\left(\sqrt{\frac{f(r_B)}{f(r_A)}}\Omega_B\right).\label{wave:packet:relation}
\label{fab}
\end{eqnarray}
From Eq. (\ref{fab}), we can see that the effect induced by the
curved spacetime  of the Earth cannot be simply corrected by a linear shift of frequencies. Therefore, it may be challenging to compensate the transformation induced by the curvature in realistic implementations.

Indeed, such a nonlinear gravitational effect is found to influence the fidelity of the quantum channel between Alice and Bob \cite{DEBT,DEBA,wangsyn}. It is always possible to decompose the mode $\bar{a}^{\prime}$ received by Bob in terms of the mode $a^{\prime}$ prepared by Alice  and an orthogonal mode $\hat{a}_{\bot}^{\prime}$ (i.e. $[\hat{a}^{\prime},\hat{a}_{\bot}^{\prime\dagger}]=0$) \cite{PPRW}
\begin{eqnarray}
\bar{a}^{\prime}=\Theta \hat{a}^{\prime}+\sqrt{1-\Theta^2}\hat{a}_{\bot}^{\prime},\label{mode:decomposition}
\end{eqnarray}
where $\Theta$ is the wave packet overlap between the distributions $F^{(B)}_{\Omega_{B,0}}(\Omega_B)$ and $F^{(A)}_{\Omega_{A,0}}(\Omega_B)$ which is given by
\begin{eqnarray}
\Theta:=\int_0^{+\infty}d\Omega_B\,F^{(B)\star}_{\Omega_{B,0}}(\Omega_B)F^{(A)}_{\Omega_{A,0}}(\Omega_B).\label{single:photon:fidelity}
\end{eqnarray}
For $\Theta=1$ corresponds to a perfect channel and the channel
between him and Alice (i.e., the spacetime) is noisy with $\Theta<1$. The
quality of the channel can be quantified by employing the
fidelity $\mathcal{F}=|\Theta|^2$.
\
Since the source is not monochromatic, we need a
frequency distribution for the mode.
We assume that Alice employs a real normalized Gaussian wave packet
\begin{eqnarray}
F_{\Omega_0}(\Omega)=\frac{1}{\sqrt[4]{2\pi\sigma^2}}e^{-\frac{(\Omega-\Omega_0)^2}{4\sigma^2}}\label{Bobpacket},
\end{eqnarray}
with  wave packet width $\sigma$. In this case the overlap $\Theta$ is given by \eqref{single:photon:fidelity} where we have extended the domain of integration to all the real axis. We note that the integral should be performed over strictly positive frequencies. This is justified since the peak frequency is
typically much larger than the spreading of the wave packet (i.e.,$\Omega_0\gg \sigma$). Thus, it is possible
to include negative frequencies without affecting the value of $\Theta$. Employing Eqs. \eqref{wave} and \eqref{Bobpacket} one finds that
\begin{eqnarray} \label{theta}
\Theta=\sqrt{\frac{2}{1+(1+\delta)^2}}\frac{1}{1+\delta}e^{-\frac{\delta^2\Omega_{B,0}^2}{4(1+(1+\delta)^2)\sigma^2}}\label{final:result},
\end{eqnarray}
where the new parameter $\delta$ quantifying the shifting is defined by
\begin{equation}
\delta=\sqrt[4]{\frac{f(r_A)}{f(r_B)}}-1=\sqrt{\frac{\Omega_B}{\Omega_A}}-1.
\end{equation}
The expression for $\frac{\Omega_B}{\Omega_A}$ in the equatorial plane of the Kerr spacetime has been shown in \cite{kerr}
\begin{equation}\label{aw}
\frac{\Omega_B}{\Omega_A}=\frac{1+\epsilon \frac{a}{r_B}\sqrt{\frac{M}{r_B}}}{C\sqrt{1-3\frac{M}{r_B}+
2\epsilon\frac{a}{r_B}\sqrt{\frac{M}{r_B}}}},
\end{equation}
where $C=[1-\frac{2M}{r_A}(1+2a {\omega})+\big(r^2_A+a^2-\frac{2Ma^2}{r_A}\big){\omega}^2]^{-\frac{1}{2}}$ is the normalization constant, $\omega$ is the Earth's equatorial angular velocity and $\epsilon=\pm1$ stand for the direct of orbits (i.e., when $\epsilon=+1$ for the satellite co-rotates with the Earth). In the Schwarzschild limit $a, \omega\rightarrow0$,  Eq. (\ref{aw}) coincides to the result found in \cite{DEBT}, which is
\begin{equation}
\frac{\Omega_B}{\Omega_A}=\sqrt{\frac{1-\frac{2M}{r_A}}{1-\frac{3M}{r_B}}}.
\end{equation}

In order to obtain the explicit expression of the frequency shift for the photon
exchanged between Alice and Bob, we expand the Eq. (\ref{aw}) and
obtain the following perturbative expression for $\delta$ by $(r_A \omega)^2>a\omega$, therefore we can retain second order terms in $r_A\omega$.  This perturbative result does not depend
on whether the Earth and the satellite are co-rotating or not
\begin{eqnarray}\label{bw}
\nonumber\delta&=&\delta_{Sch}+\delta_{rot}+\delta_h\\
\nonumber&=&\frac{1}{8}\frac{r_S}{r_A}\big(\frac{1-2\frac{h}{r_A}}{1+\frac{h}{r_A}} \big)-\frac{(r_A\omega)^2}{4}-\frac{(r_A\omega)^2}{4}\big(\frac{3}{4}\frac{r_S}{r_A}-\frac{4Ma}{\omega r_A^3}\big),
\end{eqnarray}
where $h=r_B-r_A$ is the height between Alice and Bob, $\delta_{Sch}$ is the first order Schwarzschild term, $\delta_{rot}$ is the lowest order rotation term and $\delta_h$ denotes all higher order correction terms.
If the parameter $\delta=0$ (i. e. the satellite moves at the height $h\simeq\frac{r_A}{2}$), we have $\Theta=1$.
The height at which the gravitational effect of
the Earth and the special relativistic effect (i.e., doppler effect) due to the motion of the satellite compensate each other. That is to say, the received photons by Bob at this height will not experience any frequency shift and Bob's
clock rate becomes equal to the clock rate of Alice in this height. Indeed, the satellite's motion around the Earth slows down Bob's proper time, but the higher altitude of Bob introduces a lower redshift which therefore has also a lower
effect on Bob's clock rate, as compared to Alice.
Meanwhile, the relevant limit of the expression for $\frac{\Omega_B}{\Omega_A}$ that in Minkowski is equal to $1$.

\section{Quantifying coherence of Gaussian states  \label{GSCDGE}}
%--------------------------------------------------------------------------------
In this section we briefly review the measurement of quantum coherence for a  general two-mode Gaussian state $\rho_{AB}$ which is composed of a subsystem A   and a subsystem B \cite{weedbrook}.
Then we can define the vector of the field quadratures as $\hat R = (\hat x^A,\hat p^A, \hat x^B,\hat p^B)^{\sf T}$, which satisfies the canonical commutation relations $[{{{\hat R}_k},{{\hat R}_l}} ] = i{\Omega _{kl}}$, with $\Omega = {{\ 0\ \ 1}\choose{-1\ 0}}^{\bigoplus{2}}$ being the symplectic form.
All Gaussian
properties can be determined from the symplectic form
of the covariance matrix (CM) defined as ${\sigma _{ij}} = \text{Tr}\big[ {{{\{ {{{\hat R}_i},{{\hat R}_j}} \}}_ + }\ {\rho _{AB}}} \big]$  \cite{RSP,RSP1,RSP2,RSP3}
\begin{eqnarray}\label{wsm11}
 \sigma_{AB}= \left(\!\!\begin{array}{cccc}
a&0&c_1&0\\
0&a&0&-c_2\\
c_1&0&b&0\\
0&-c_2&0&b
\end{array}\!\!\right).
\end{eqnarray}
The correlations $a$,  $b$,  $c_1$ and $c_2$ are determined by the four local symplectic invariants $I_1=a^2$, $I_2=b^2$, $I_3=c_1c_2$ and $I_4=\det(\sigma_{AB})=(ab-c_1^2)(ab-c_2^2)$.
The symplectic eigenvalues of the CM of a two-mode Gaussian state are given as
$2{\nu}_{\mp}^2=\Delta \mp\sqrt{\Delta^2-4I_4}$ with $\Delta=I_1+I_2+2I_3$ \cite{RSP2,RSP3}.

The coherence measure $C(\rho _{AB})$ has been given in terms of the
displacement vectors and covariance matrix in \cite{JWX}. Then we use the coherence measure as
$C(\rho _{AB})=\inf{S(\rho _{AB}||\delta_{AB})}$,  where $\delta_{AB}$ is the nearest incoherent Gaussian state of $\rho _{AB}$. The von Neumann entropy of a bipartite system $\rho _{AB}$ in terms of the symplectic eigenvalues is given by \cite{RSP5}

 \begin{eqnarray}\label{wsm12}
 S(\rho _{AB})=f({\nu}_-)+f({\nu}_+),
 \end{eqnarray}
where $f(\nu)=\frac{\nu+1}{2}\log_2\frac{\nu+1}{2}-\frac{\nu-1}{2}\log_2\frac{\nu-1}{2}$, while the mean occupation value is \cite{JWX}

\begin{eqnarray}\label{wsm13}
 \overline{n}_k=\frac{1}{4}(\sigma_{11}^k+\sigma_{22}^k+[d_1^k]^2+[d_2^k]^2).
 \end{eqnarray}
Here, $\sigma^1$ and $\sigma^2$ are elements of the subsystem of A and B in CM, respectively, and $[d^k_i]^2$ is $i$ first statistical moment of the $k$ mode. For convenience, we
select $d^k_1=d^k_2=0$.
It is possible to obtain an analytical expression of the quantum coherence of Gaussian states \cite{JWX}
\begin{eqnarray}\label{wsm14}
\nonumber C({\rho _{AB}})=&-&S(\rho _{AB})+\Sigma_{i=1}^2[(\overline{n}_i+1)\log_2(\overline{n}_i+1)\\
&-&\overline{n}_i\log_2\overline{n}_i].
\end{eqnarray}

\section{The influence of gravitational effect on Gaussian coherence}

In this section we propose a scheme to test large distance quantum coherence between two  satellites with different heights and discuss how quantum coherence is affected by the curved spacetime of the Earth. Firstly, we consider a pair of entangled photons
which are initially prepared in a two-mode squeezed state with modes $b_1$ and $b_2$ at the ground station. Then we send one photon with mode  $b_1$ to Alice. The other photon in mode $b_2$ propagates from the
Earth to the satellite and is received by Bob (at the height $h_B=r_B-r_A$). Due to the curved background spacetime of the Earth, the wave packet of photons are deformed. Finally, we study the behavior of Gaussian coherence under the Earth's gravitational field.

Considering that Alice receives the mode $b_1$ and Bob receives the mode $b_2$ at different satellite orbits, we should take the curved spacetime  of the Earth into account. As discussed in \cite{DEBT,DEBA,wangsyn}, the influence of the Earth's  gravitational  effect can be modeled by a beam splitter with orthogonal modes $b_{1\bot}$ and $b_{2\bot}$. The covariance matrix of the initial state is given by
\begin{equation}\label{initialcov}
\Sigma^{b_1b_2b_{1\bot}b_{2\bot}}_0=\left(
\begin{array}{cc} \tilde\sigma(s) &0 \\ 0  &  I_4
\end{array}\right),
\end{equation}
where ${I}_4$ denotes the $4\times4$ identity matrix and $\tilde\sigma{(s)}$ is  the covariance matrix of the two-mode squeezed state
\begin{equation}
\tilde\sigma(s)=\left(
\begin{array}{cc} \cosh{(2s)}  {I}_2&\sinh{(2s)}\sigma_z \\ \sinh(2s)\sigma_z &\cosh{(2s)}  {I} _2
\end{array}\right),
\end{equation}
where $\sigma_z$ is Pauli matrix and $s$ is  the squeezing parameter. The effects induced by the
curved spacetime of the Earth on Alice's  mode $b_1$ and Bob's mode $b_2$ can be model as lossy channel, which are described by the transformation \cite{DEBT,DEBA,wangsyn}
\begin{eqnarray}
\bar{b}_1&=&\Theta_1\,b_1+\sqrt{{1-\Theta_1^2}}b_{1\bot},\\
\bar{b}_2&=&\Theta_2\,b_2+\sqrt{{1-\Theta_2^2}}b_{2\bot}.
\end{eqnarray}
This process can be represented as a mixing (beam splitting ) of modes $b_1(b_2)$ and $b_{1\bot}(b_{2\bot})$. Therefore, for the entire state, the symplectic transformation can be encoded into the Bogoloiubov transformation
\begin{equation}
S=\left(
\begin{array}{cccc}
  \Theta_1{I}_2 &0&  \sqrt{1-\Theta_1^2}&0  \\ 0&\Theta_2  {I} _2 &0&\sqrt{
1-\Theta_2^2}  {I} _2\\  \sqrt{1-\Theta_1^2} &0&  -\Theta_1{I}_2&0  \\ 0&\sqrt{
1-\Theta_2^2}  {I} _2 &0&-\Theta_2  {I} _2
\end{array}\right).\nonumber
\end{equation}
The final state  $\Sigma^{b_1b_2b_{1\bot}b_{2\bot}}$ after the transformation is  $\Sigma^{b_1b_2b_{1\bot}b_{2\bot}}=S\,\Sigma_0^{b_1b_2b_{1\bot}b_{2\bot}}\,S^{T}$.  Then we trace over the orthogonal modes $b_{1\bot},b_{2\bot}$ and obtain the covariance matrix $\Sigma^{b_1b_2}$ for the modes $b_1$ and $b_2$  after the propagation
\begin{equation}\Sigma^{b_1b_2}=\left(
\begin{array}{cc}\label{gst}
(1+2\sinh^2s \Theta_1^2) {I}_2 &\sinh{(2s)}\,\Theta_1\Theta_2\,\sigma_z  \\ \sinh{(2s)}\,\Theta_1\Theta_2\,\sigma_z &(1+2\sinh^2s\,\Theta_2^2 )\, {I}_2
\end{array}\right).
\end{equation}

The form of the two-mode squeezed state under the influence of the effects of gravity of the Earth is given by Eq. (\ref{gst}). Then employing Eq . (\ref{wsm14}), we can obtain Gaussian coherence between the mode $b_1$ and $ b_2$ under the curved spacetime of the Earth. We notice that the effect of the Earth on the quantum state of the photon is modeled by a lossy quantum channel which is determined by the wave packet overlap parameter $\Theta$ that contains parameters $\delta$, $\sigma$ and $\Omega_{B,0}$.  Since the Schwarzschild radius of the Earth is $ r_S= 9$ mm, and we constrain the satellite height to geostationary Earth orbits, we have
$\delta\sim 2.5\times10^{-10}$.  Here we consider a typical parametric down converter crystal (PDC) source with a
wavelength of 598 nm (corresponding to the peak frequency $\Omega_{B,0}= 500$ THz) and   Gaussian bandwidth $\sigma=1$MHz \cite{NCMS, DNMP}. Under these constraints, $\delta\ll(\frac{\Omega_{B,0}}{\sigma})^2\ll1$ is
satisfied.
Therefore, the wave packet overlap $\Theta$ can be expand by the parameter $\delta$.  Then we obtain $\Theta \sim1-\frac{\delta^2\Omega_{B,0}^2}{8\sigma^2}$ by keeping the second order terms.

 \iffalse To ensure the validity of  perturbative expansion, we  estimate the values  $\frac{\delta^2\Omega^2_{B,0}}{2\sigma^2}\sim1.25\times10^{-7}$, and find that even if the value of the squeezing parameter is $s\ll7.6$ (corresponding to $\sinh^2(s)\ll10^6$), the perturbative expansion is valid. Therefore, we can safely prelimit the value of the squeezing parameter as $s<3$ hereafter.
 \fi

\begin{figure}[tbp]

\centering
\centerline{\includegraphics[width=9.2cm]{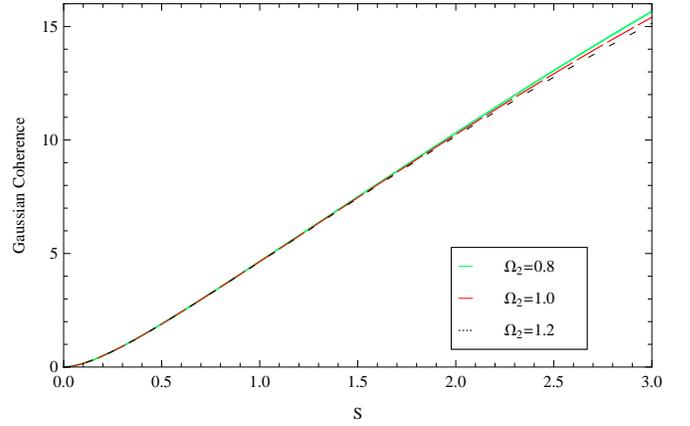}}
\caption{(Color online) The Gaussian coherence $C(\Sigma^{b_1 b_2})$ as a function of the squeezing parameter $s$ for different peak frequencies, $\Omega_2=0.8$ (green solid line), $\Omega_2=1$ (red dashed line) and $\Omega_2=1.2$ (black dashed line), respectively. The orbit height of the satellite and the Gaussian bandwidth are fixed as $h=20000$ km and  $\sigma=1$.
}\label{f1}
\end{figure}

For convenience, we will work with dimensionless quantities by rescaling the peak frequency and the Gaussian bandwidth
\begin{equation}
\Omega \rightarrow \tilde{\Omega}\equiv\frac{\Omega}{\Omega_{B,0}}, \sigma \rightarrow \tilde{\sigma}\equiv\frac{\sigma}{\sigma_0},
\end{equation}
where $\Omega_{B,0}=500$THz and $\sigma_0=1$ MHz. For simplicity, we abbreviate  the dimensionless parameter $\tilde{\Omega}$ as $\Omega_2$ and abbreviate $\tilde{\sigma}$ as  $\sigma$, respectively.

To better understand the relation between Gaussian coherence and initial squeezing parameter.
In Fig. (1) we plot the Gaussian coherence $C(\Sigma^{b_1 b_2})$ as a function of the squeezing parameter $s$
for the fixed orbit height $h=20000$ km and Gaussian bandwidth $\sigma=1$. We can see that Gaussian coherence monotonically increases with the increase of the squeezing parameter $s$. We also can see that Gaussian coherence decreases with the growth of the peak frequency parameter of the mode $b_2$. However, comparing with the peak frequency parameter, Gaussian coherence $C(\Sigma^{b_1 b_2})$ is easier effected by  changing  squeezing parameters.  That is to say, the Gaussian coherence is more sensitive to squeezing parameter than peak frequency parameter.

\begin{figure}[tbp]
\centering
\includegraphics[height=2.2in, width=3.2in]{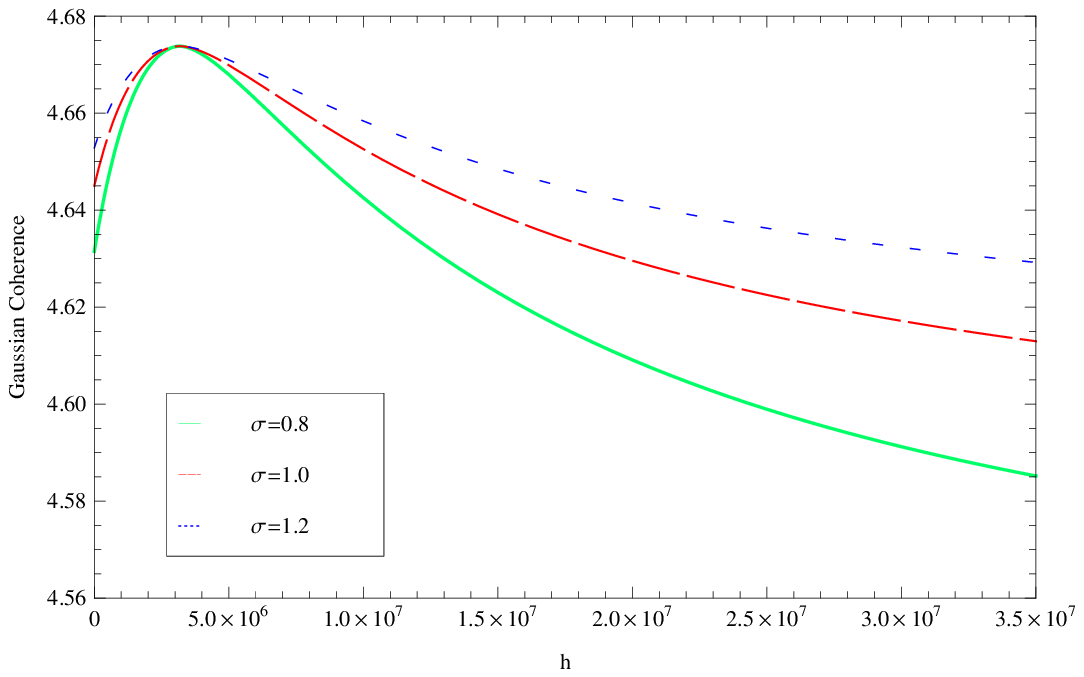}
\caption{(Color online) The Gaussian coherence $C(\Sigma^{b_1 b_2})$ in terms of the orbit height $h$ for different values of Gaussian bandwidth, $\sigma=0.8$ (green solid line), $\sigma=1$ (red dashed line) and $\sigma=1.2$ (blue dashed line), respectively. The squeezing parameter and peak frequency of mode $b_2$ are fixed as $s=1$ and  $\Omega_2=1$. }
\end{figure}

\begin{figure}[tbp]
\centerline{\includegraphics[height=2.2in, width=3.2in]{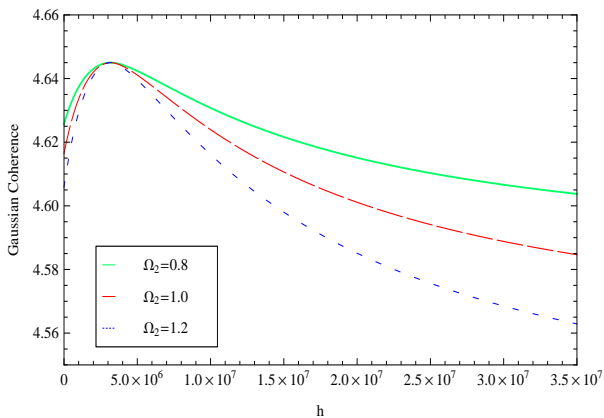}}
\caption{ (Color online). The Gaussian coherence $C(\Sigma^{b_1 b_2})$ in terms of the orbit height $h$ for different peak frequencies of mode $b_2$, $\Omega_2=0.8$ (green solid line), $\Omega_2=1$ (red dashed line) and $\Omega_2=1.2$ (blue dashed line), respectively. The other parameters are fixed as $s=1$  and  $\sigma=1$.
}\label{f1}
\end{figure}

The behavior of Gaussian coherence under the Earth's gravitational field has been shown in Fig. (2) and (3).
The Gaussian coherence $C(\Sigma^{b_1 b_2})$ in terms of the orbit height $h$ with the different values of the Gaussian bandwidth has been shown in Fig. (2). Meanwhile, we plot the $C(\Sigma^{b_1 b_2})$ in terms of the orbit height $h$ for different peak frequencies of mode $b_2$ in Fig. (3). Comparing these two pictures, we can see that the Gaussian coherence increases with the increase of Gaussian bandwidth parameter, but the Gaussian coherence decreases with the growth of the peak frequency.
Moreover, the typical distance between the Earth and the geostationary
satellite is about $3.6\times10^4$ km, which yields the height $r_B = 4.237\times10^4$ km for the satellite. For this distance, the influence of relativistic disturbance of the spacetime curvature on quantum coherence cannot be ignored for the quantum information tasks at current level technology \cite{satellite1, satellite2, MJAG}. Hence, we constrain the satellite height to geostationary Earth orbits $r_B(GEO) = r_A+3.6\times10^4$ km.

The Fig. (2) and Fig. (3) both shown that Gaussian coherence increases for a specific range of height parameter $h\simeq\frac{r_A}{2}$ and then gradually approach to a finite value with increasing $h$. The physical support behind this is that the
gravitational frequency shift effects would reduce quantum resource, but the special
relativistic effects makes quantum resource growth. Since the special
relativistic effects becomes smaller and smaller but the gravitational frequency shift can be cumulate with increasing height. The photon's frequency received by satellites with height $h<\frac{r_A}{2}$ will experience  blue-shift which cause Gaussian coherence increases, while the frequencies of photons received at height $h>\frac{r_A}{2}$ experience red-shift which cause Gaussian coherence decreases.
In fact, the peak value of Gaussian coherence (the parameter $\delta=0$) indicates the fact that the photon's frequency received by satellites experiences a transformation from blue-shift to red-shift, which causes the Gaussian coherence between the photon pairs to increase first and then to reduce with increasing height \cite{kerr}.
When two parties are situated at the same height or are in flat space-time, the parameter $\delta\neq 0$. It comes from the fact that we are expanding the
total frequency shift in Eq. (\ref{aw}) taking into account both special
and general relativistic effects \cite{kerr}. When the satellite moves at the height $h=\frac{r_A}{2}$, the Schwarzschild term $\delta_{Sch}$ vanishes and photons received on satellites will generate a very small frequency shift effects, therefore the lowest order rotation term $\delta_{rot}$ needs to be considered. In addition, Gaussian coherence is not equal with different $\Omega_2$ and $\sigma$ when Alice and Bob at  height $h=0$. The reason for this result is that $\delta\neq0$ when height $h=0$, the contribution of the special relativity effects always existence which leads to $\Theta$ parameter is not equivalent to zero. And the parameter $\Theta$ not only depends on satellite's height but also depends on  Gaussian bandwidth parameter and  peak frequency which means that different Gaussian bandwidth parameters and  peak frequencies correspond to different $\Theta$ parameters.

\begin{figure}[tbp]
\centering
\includegraphics[height=2.3in, width=3.3in]{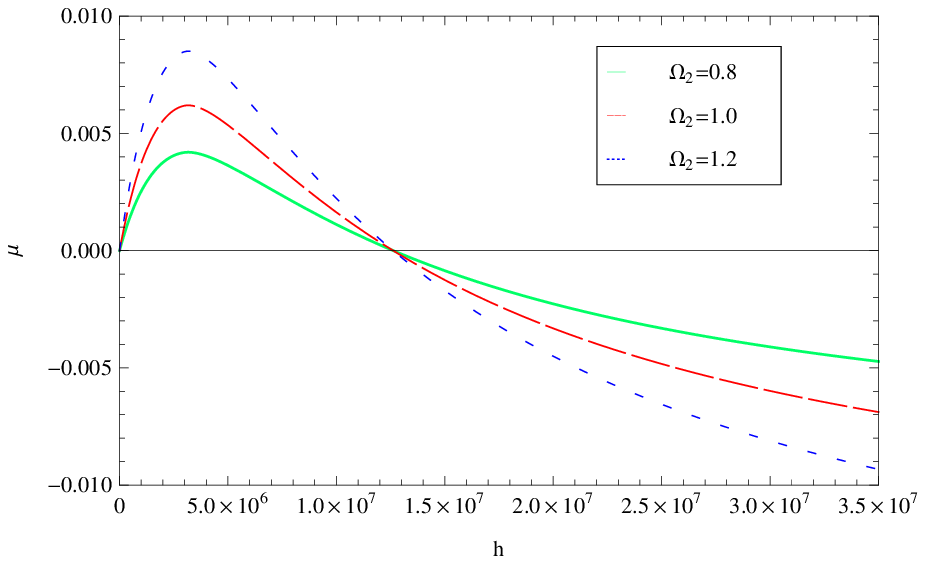}
\caption{(Color online) The Gaussian coherence $C(\Sigma^{b_1 b_2})$ in terms of the orbit height $h$ for different values of Gaussian bandwidth, $\sigma=0.8$ (green solid line), $\sigma=1$ (red dashed line) and $\sigma=1.2$ (blue dashed line), respectively. The squeezing parameter and peak frequency of mode $b_2$ are fixed as $s=1$ and  $\Omega_2=1$. }
\end{figure}

Consider Gaussian coherence is a quantum resource, understanding how much Gaussian coherence affected by the Earth's gravitational field is more important. We calculate the rate of change of Gaussian coherence $\mu$ according follow equation

\begin{equation}
\mu=\frac{C(\Sigma^{b_1b_2})-C_0(\Sigma^{b_1b_2})}{C_0(\Sigma^{b_1b_2})},
\end{equation}
where $C_0(\Sigma^{b_1b_2})$ is value of Gaussian coherence $C(\Sigma^{b_1b_2})$ with the satellite's height $h=0$, the parameter $\mu$ means the degree of change of Gaussian coherence suffers  the Earth's gravitational field. We shown the rate of change of Gaussian  coherence $\mu$ for different peak frequencies $\Omega_2$ with the fixed $\sigma=1$ in Fig. (4). It is easy to see that value of $\mu$ increases when height parameter $h<\frac{r_A}{2}$, then gradually decreases, which indicates that the Earth's gravitational effect for photon's frequency shift is blue-shift in $h<\frac{r_A}{2}$ range, for $h>\frac{r_A}{2}$ range, is red shift. It is also shown that total gravitational frequency shift causes the change of Gaussian coherence less than $\% 1$  with the satellite height in
geostationary Earth orbits.  In addition, the point $h=2r_A$ (i.e. the parameter
$\delta=-1$) means that the Earth's gravitational frequency shift for photon is no contribution, the blue-shift and the red-shift effect for photon's frequency are offset, there's only special relativity effect. The different initial peak frequencies also can effect rate of change $\mu$. This conclusion give us guide to choose appropriate physical parameters to constrains unnecessary loss of Gaussian coherence between the Earth to satellites.

\vspace*{0.8cm}

\section{Conclusions}

In conclusion, we have studied Gaussian coherence for a two-mode Gaussian state when one of the modes propagates from the ground to satellites.
We found that the frequency shift induced by the curved spacetime  of the Earth reduces the quantum correlation of coherence between the photon pairs  when  one of the entangled photons is sent to the Earth station and the other photon is sent  to the satellite.
We also found that Gaussian coherence is easier to change
with the initial squeezing parameter than the gravitational effect and other parameters.  Meanwhile, we found that Gaussian coherence increases with the increase of Gaussian bandwidth parameter, but the Gaussian coherence decreases with the growth of the peak frequency.
In addition, the peak value is found to be a critical point which indicates the Gaussian coherence experiences the blue-shift transforms into the
red-shift. Finally, it is also found that total gravitational frequency shift causes the change of Gaussian coherence less than $\% 1$ and  different initial peak frequencies also can effect rate of change with the satellite height in geostationary Earth orbits.
According to the equivalence principle, the effects of acceleration are equivalence with the effects of gravity, our results could be in principle apply to dynamics of quantum coherence under the influence of acceleration. Since realistic quantum systems will always exhibit gravitational
and relativistic features, our results should be significant both for giving more advices to realize quantum information protocols such
as quantum key distribution from Earth to satellites and for a general understanding of quantum coherence in  relativistic quantum systems.

\section*{Acknowledgments}

\begin{acknowledgments}

This work was supported by National Key R\&D Program of China No.
2017YFA0402600; the National Natural Science Foundation of China
under Grants Nos. 11690023, and 11633001; Beijing Talents Fund of
Organization Department of Beijing Municipal Committee of the CPC;
the Fundamental Research Funds for the Central Universities and
Scientific Research Foundation of Beijing Normal University; and the
Opening Project of Key Laboratory of Computational Astrophysics,
National Astronomical Observatories, Chinese Academy of Sciences.	
\end{acknowledgments}


\begin{thebibliography}{99}
\bibitem{E.P.C}
A. J. Leggett, Prog. Theor. Phys. Suppl. {\bf 69}, 80 (1980).

\bibitem{EPC1}
V. Giovannetti, S. Lloyd, and L. Maccone, Science. {\bf 306}, 1330 (2004).

\bibitem{EPC2}
R. Demkowicz-Dobrzanski and L. Maccone, Phys. Rev. Lett. {\bf 113}, 250801 (2014).

\bibitem{EPC3}
J. K. Asb¨®th, J. Calsamiglia, and H. Ritsch, Phys. Rev. Lett. {\bf 94}, 173602 (2005).

\bibitem{EPC4}
A. Streltsov, U. Singh, H. S. Dhar, M. N. Bera, and
G. Adesso, Phys. Rev. Lett. {\bf 115}, 020403 (2015).

\bibitem{EPC5}
R. J. Glauber, Phys. Rev. {\bf 131}, 2766 (1963).

\bibitem{EPC6}
M. O. Scully, Phys. Rev. Lett. {\bf 67}, 1855 (1991).

\bibitem{EPC7}
A. Albrecht, J. Mod. Opt. {\bf 41}, 2467 (1994).

\bibitem{EPC8}
D. F. Walls and G. J. Milburn, \emph{Quantum Optics} (Springer-
Verlag, Berlin, 1995).

\bibitem{EPC9}
M. Nielsen and I. Chuang, \emph{Quantum Computation
and Quantum Information} (Cambridge University Press, Cambridge, England, 2000), ISBN: 9781139495486.

\bibitem{EPC10}
T. Baumgratz, M. Cramer, and M. B. Plenio, Phys. Rev.
Lett. {\bf 113}, 140401 (2014).

\bibitem{EPC11}
F. Levi and F. Mintert, New J. Phys. {\bf 16}, 033007 (2014).

\bibitem{EPC12}
I. Marvian and R.W. Spekkens, New J. Phys. {\bf 15}, 033001 (2013).

\bibitem{EPC13}
J. {\AA}berg, arXiv:quant-ph/0612146 (2006).

%\bibitem{EPC14}
 %D. Girolami, Phys. Rev. Lett. 113, 170401.


\bibitem{HMW}
S. D. Bartlett, T. Rudolph, and R.W. Spekkens, Rev. Mod.
Phys. {\bf 79}, 555 (2007).


\bibitem{Skrzypczyk}
I. Marvian, Ph.D. thesis, University of Waterloo, 2012.


\bibitem{Kocsis}
D. Girolami, T. Tufarelli, and G. Adesso, Phys. Rev. Lett.
{\bf 110}, 240402 (2013).

\bibitem{Adesso2015}
I.Marvian and R.W. Spekkens, Nat.Commun. {\bf 5}, 3821 (2014)


\bibitem{steering3}
D. Girolami, Phys. Rev. Lett. {\bf 113}, 170401 (2014).

\bibitem{steering4}
D. Girolami, A. M. Souza, V. Giovannetti, T. Tufarelli, J. G.
Filgueiras, R. S. Sarthour, D. O. Soares-Pinto, I. S. Oliveira,
and G. Adesso, Phys. Rev. Lett. {\bf 112}, 210401 (2014).

\bibitem{MWZZ}
J. Aberg, Phys. Rev. Lett. {\bf 113}, 150402 (2014).

\bibitem{steering5}
S. Luo, S. Fu, and C. H. Oh, Phys. Rev. A. {\bf 85}, 032117 (2012).


\bibitem{Walborn}
X. Yuan, H. Zhou, Z. Cao, and X. Ma, Phys. Rev. A. {\bf 92},
022124 (2015).


\bibitem{steering2}
Z. Xi, Y. Li, and H. Fan, Sci. Rep. {\bf 5}, 10922 (2015).

\bibitem{Bowles}
A.Winter and D. Yang, Phys. Rev. Lett. {\bf 116}, 120404 (2016).

\bibitem{VHTE}
E. Chitambar, A. Streltsov, S. Rana, M. N. Bera, G. Adesso,
and M. Lewenstein, Phys. Rev. Lett. {\bf 116}, 070402 (2016).

\bibitem{DSSC}
D. Samos-S\`{a}enz D. Buruaga, C. Sab, Phys. Rev. A. {\bf 95}, 022307 (2017).


\bibitem{TBM}
J. Wang, Z. Tian, J. Jing and H. Fan, Phys. Rev. A {\bf93}, 062105 (2016).


\bibitem{Alclock}
C. Chou, D. Hume, T. Rosenband,
and D. Wineland,  Science.  {\bf 329}, 1630 (2010).

\bibitem{wangsyn}
J. Wang, Z. Tian, J. Jing, and H. Fan, Phys. Rev. D. {\bf 93}, 065008 (2016).

\bibitem{Zych}
M. Zych, F. Costa, I. Pikovski, and C. Brukner, Nat. Commun. {\bf  2}, 505 (2011).


\bibitem{MADE}
M. Ahmadi, D. Bruschi, and I. Fuentes, Phys. Rev. D. {\bf  89}, 065028 (2014).

\bibitem{MADE2}
M. Ahmadi, D. Bruschi, C. Sab\`{\i}n, G. Adesso, and
I. Fuentes, Sci. Rep. {\bf 4}, 4996 (2014).


\bibitem{SPK}
S. Kish and T. Ralph, Phys. Rev. D. {\bf 93}, 105013 (2016).

\bibitem{TQJJH}
T. Liu, J. Jing, J. Wang, Adv. Q. T. {\bf 1}, 2 (2018).


\iffalse
\bibitem{RQI8}
M. Fink, A. Rodriguez-Aramendia, J. Handsteiner, A. Ziarkash, F. Steinlechner, T. Scheidl, I. Fuentes, J. Pienaar, T, Ralph, and R. Ursin, Nat. Commun.  {\bf 8}, 15304 (2017).\fi

\bibitem{MANI}
M. Nielsen and I. Chuang,
\emph{Quantum computation and quantum information} (Cambridge
University Press, 2000).

\bibitem{PPRW}
P. Rohde, W. Mauerer, and C. Silberhorn, New J. Phys. {\bf 9}, 91 (2007).


\bibitem{DEBT}
D. Bruschi, T. Ralph, I. Fuentes, T. Jennewein,
and M. Razavi, Phys. Rev. D. {\bf  90}, 045041 (2014).

\bibitem{Visser}
M. Visser, arXiv:0706.0622 (2007).

\bibitem{ULMQ}
U. Leonhardt, \emph{Measuring the Quantum State of Light,
Cambridge Studies in Modern Optics} (Cambridge University
Press, Cambridge, 2005).

\bibitem{TGDT}
T. Downes, T. Ralph, and N. Walk, Phys. Rev. A.
{\bf  87}, 012327 (2013).

\bibitem{DEBA}
D. Bruschi, A. Datta, R. Ursin, T. Ralph, and I. Fuentes,
Phys. Rev. D. {\bf 90}, 124001 (2014).

\bibitem{weedbrook}
C. Weedbrook, S. Pirandola, R. Garc\'ia-Patr\'on, N. J. Cerf, T. C.
Ralph, J. H. Shapiro, and S. Lloyd, Rev. Mod. Phys. {\bf 84}, 621
(2012).

\bibitem{RSP}
R. Simon, Phys. Rev. Lett. {\bf 84}, 2726 (2000).

\bibitem{RSP1}
L.-M. Duan, G. Giedke, J. I. Cirac, and P. Zoller, Phys. Rev.
Lett. {\bf84}, 2722 (2000).

\bibitem{RSP2}
D. Buono, G. Nocerino, A. Porzio, and S. Solimeno, Phys.
Rev. A. {\bf 86}, 042308 (2012).

\bibitem{RSP3}
F. A. S. Barbosa, A. J. de Faria, A. S. Coelho, K. N.
Cassemiro, A. S. Villar, P. Nussenzveig, and M. Martinelli,
Phys. Rev. A. {\bf 84}, 052330 (2011).


\bibitem{JWX}
J. W. Xu, Phys.Rev. A. {\bf 93}, 032111 (2016).

\bibitem{RSP5}
A. S. Holevo, M. Sohma, and O. Hirota, Phys. Rev. A. {\bf 59}, 1820
(1999).

\bibitem{NCMS}
M. Razavi and J. Shapiro, Phys. Rev. A. {\bf73},
042303 (2006).

\bibitem{DNMP}
D. Matsukevich, P. Maunz, D. Moehring, S. Olmschenk,
and C. Monroe, Phys. Rev. Lett. {\bf 100},
150404 (2008).

\bibitem{satellite1}
G. Vallone, D. Bacco, D. Dequal, S. Gaiarin, V. Luceri, G. Bianco, and P. Villoresi, Phys. Rev. Lett. {\bf  115}, 040502 (2015).

\bibitem{satellite2}
J. Yin \emph{et. al}, Science.  {\bf 356}, 1140 (2017).

\bibitem{MJAG}
M. Jofre, A. Gardelein, G. Anzolin, W. Amaya, J. Capmany,
R. Ursin, L. Penate, D. Lopez, J. Juan, J.
Carrasco, Opt. Express {\bf 19}, 3825 (2011).

\bibitem{kerr}
J. Kohlrus, D. Bruschi, J. Louko,  and I. Fuentes,
EPJ Quantum Technology. {\bf  4}, 7 (2017).


\iffalse
\bibitem{VHTSS}
S. Wollmann, N. Walk, A. Bennet, H. Wiseman, and G. Pryde,
\emph{Phys. Rev. Lett {\bf 2016}, 116}, 160403.


\bibitem{Handchen}
T. Guerreiro, F. Monteiro, A. Martin, J. Brask, T. V\'{e}rtesi, B. Korzh, M. Caloz, F. Bussi$\grave{ e}$res, V. Verma, A. Lita, R. Mirin, S. Nam, F. Marsilli, M. Shaw, N. Gisin, N. Brunner, H. Zbinden, and R. Thew, \emph{Phys. Rev. Lett {\bf 2016}, 117}, 070404.


\bibitem{BWSR}
B. Wittmann, S. Ramelow, F. Steinlechner, N. Langford,
N. Brunner, H. Wiseman, R. Ursin, and A. Zeilinger,
\emph{New Journal of Physics {\bf 2012}, 14}, 053030.

\bibitem{SKMJ}
S. Kocsis, M. Hall, A. Bennet, and G. Pryde,
\emph{Nat. Commun {\bf 2015}, 6}, 5886.

\bibitem{DJSS}
D. Saunders, S. Jones, H. Wiseman, and G. Pryde,
\emph{Nat. Phys {\bf 2010}, 6}, 845.

\bibitem{TEVH}
T. Eberle, V. H$\ddot{a}$ndchen, J. Duhme, T. Franz, R. F-Werner, and R. Schnabel, \emph{Phys. Rev. A {\bf 2011}, 83}, 052329.


\bibitem{SWNW}
S. Wollmann, N. Walk, A. Bennet, H. Wiseman, and G. Pryde,
\emph{Phys. Rev. Lett {\bf 2016}, 116}, 160403.


\bibitem{steeringrqi}
M. Navascues, D. Perez-Garcia, \emph{Phys. Rev. Lett {\bf 2012}, 19}, 160405.

\bibitem{steeringrqi1}
C. Sab\'in, and G. Adesso, \emph{Phys. Rev. A  {\bf 2015}, 92}, 042107.

\bibitem{steeringrqi2}
J. Wang, H. Cao, J. Jing, and H. Fan, \emph{Phys. Rev. D {\bf 2016}, 93}, 125011.

\bibitem{steeringrqi3}
W. Sun, D. Wang, L. Ye, \emph{L. Phys. Lett {\bf 2017}, 14}, 9.








\bibitem{Alclock}
C. Chou, D. Hume, T. Rosenband,
and D. Wineland,  \emph{Science  {\bf 2010}, 329}, 1630.


\bibitem{Zych}
M. Zych, F. Costa, I. Pikovski, and C. Brukner, \emph{Nat. Commun {\bf 2011}, 2}, 505.



\bibitem{SPK}
S. Kish and T. Ralph, \emph{Phys. Rev D {\bf 2016}, 93}, 105013.

%\bibitem{TZNG}
%T. van Zoest, N. Gaaloul, Y. Singh, H. Ahlers, W. Herr,
%S. T. Seidel, W. Ertmer, E. Rasel, M. Eckart, E. Kajari,
%et al., Science {\bf 328}, 1540 (2010).
%
%\bibitem{CSDB}
%C. Sab\'in, D. Bruschi, M. Ahmadi, and I. Fuentes, New
%J. Phys. {\bf 15}, 085003 (2014).
%
%\bibitem{DBCA}
%D. Bruschi, C. Sab\'in, A. White, V. Baccetti, D. Oi, and
%I. Fuentes, New J. Phys. {\bf 16}, 053041 (2014).

\bibitem{wangsyn}
J. Wang, Z. Tian, J. Jing, and H. Fan,  \emph{Phys. Rev. D {\bf 2016}, 93}, 065008.

\bibitem{RQI8}
M. Fink, A. Rodriguez-Aramendia, J. Handsteiner, A. Ziarkash, F. Steinlechner, T. Scheidl, I. Fuentes, J. Pienaar, T, Ralph, and R. Ursin, \emph{Nat. Commun  {\bf2017}, 8}, 15304.

\bibitem{MANI}
M. Nielsen and I. Chuang,\emph{
Quantum computation and quantum information} (Cambridge
University Press, 2000).





%\bibitem{CWMK}
%C. W. Misner, K. S. Thorne, and J. A. Wheeler, Gravitation
%(W. H. Freeman and Company, San Francisco,1973).

\bibitem{RMWG}
R. Wald, \emph{General relativity} (The University of
Chicago Press, Chicago and London, 1984).







\bibitem{IKAR}
I. Kogias, A. Lee, S. Ragy, and G. Adesso,
\emph{Phys. Rev. Lett {\bf 2015}, 114}, 060403.

\bibitem{NCMS}
M. Razavi and J. Shapiro, \emph{Phys. Rev. A {\bf2006}, 73},
042303.

\bibitem{DNMP}
D. Matsukevich, P. Maunz, D. Moehring, S. Olmschenk,
and C. Monroe, \emph{Phys. Rev. Lett. {\bf 2008}, 100},
150404.


\bibitem{HMWS}
H. Wiseman, S. Jones, and A. Doherty, \emph{Phys. Rev. Lett
{\bf 2007}, 98}, 140402.

\bibitem{renyi}
G. Adesso, D. Girolami, and A. Serafini, \emph{Phys. Rev. Lett {\bf 2012}, 109},
190502.


\bibitem{MAK}
M. Ahmadi, K. Lorek, A. Ch\c{e}ci\'{n}ska, H. Smith, R. Mann, A. Dragan,
\emph{Phys. Rev. D {\bf 2016}, 93}, 124031.
\fi
\end{thebibliography}
\end{document}